\documentclass{journal_style}
\usepackage{url} 
\usepackage{soul}
\usepackage{fix-cm}
\usepackage{xurl}
\journalname{Water Resources Research}
\usepackage{setspace}
\begin{document}

\title{Rate dependency of capillary heterogeneity trapping for CO\textsubscript{2} storage}

\authors{Catrin Harris$^{1}$, Samuel Krevor$^{1}$, Ann H. Muggeridge$^{1}$ \& Samuel J. Jackson$^{2*}$ }

\affiliation{1}{Department of Earth Science and Engineering, Imperial College London, London, UK}
\affiliation{2}{CSIRO Energy, Clayton South, Victoria, Australia}

\correspondingauthor{Samuel J. Jackson}{samuel.jackson@csiro.au}

%%%%%%%%%%%%%%%%%%%%%%%%%%%%%%%%%%%%%%%%%%%%%%%
% KEY POINTS
%%%%%%%%%%%%%%%%%%%%%%%%%%%%%%%%%%%%%%%%%%%%%%%
%  List up to three key points (at least one is required)
%  Key Points summarize the main points and conclusions of the article
%  Each must be 140 characters or fewer with no special characters or punctuation and must be complete sentences

% Example:
% \begin{keypoints}
% \item	List up to three key points (at least one is required)
% \item	Key Points summarize the main points and conclusions of the article
% \item	Each must be 140 characters or fewer with no special characters or punctuation and must be complete sentences
% \end{keypoints}

\begin{keypoints}
\item Three-dimensional capillary heterogeneity trapping quantified in a Bentheimer sandstone at varying flow rates using medical X-Ray CT

\item Continuum analytical model of capillary heterogeneity trapping developed which agrees with experimental data

\item Trapping length and capillary heterogeneity trapping amount can be estimated using the analytical model with petrophysical properties 

\end{keypoints}

\section*{\centering Abstract}
In this paper, we experimentally quantify and analytically model rate dependent capillary heterogeneity trapping. Capillary heterogeneity trapping enhances non-wetting fluid trapping beyond pore-scale residual trapping through the isolation of non-wetting phase upstream of heterogeneities in the continuum capillary pressure characteristics. Whilst residual trapping is largely insensitive to the range of flow regimes prevalent in engineered reservoir settings, continuum theory anticipates that capillary heterogeneity trapping will be more sensitive to the balance of viscous and capillary forces that occur.
We perform steady-state drainage and imbibition multiphase flow experiments at varying flow rate on a layered Bentheimer sample with {\em in situ} medical X-ray CT scanning to quantify saturation. Saturation discontinuities are observed upstream of capillary pressure barriers as a result of capillary pressure discontinuities, trapping the non-wetting phase at a saturation greater than pore-scale residual trapping alone. We confirm the flow rate dependence predicted by theory whereby the relationship between the initial and residual saturations approach a 1:1 dependence as flow rate is decreased. We develop a one-dimensional analytical model to quantify the proportion of capillary heterogeneity trapping in the system and the dimensionless trapping length scale, which agrees with the experimental data and allows for rapid estimates of trapping up to the field-scale. 

\section*{\centering Plain Language Summary}
%The trapping of injected carbon dioxide (CO$_2$) is crucial for secure carbon sequestration in subsurface aquifers. Injected CO$_2$ is trapped most prominently by structural trapping beneath an impermeable caprock and by residual trapping whereby native aquifer water naturally imbibes the injected CO$_2$ and traps it in disconnected ganglia that cannot migrate. Aquifers generally have variations in rock types and structure that cause variations in the migration and build-up of CO$_2$. This can cause increased trapping beyond residual trapping, so-called capillary heterogeneity trapping. This trapping generally enhances storage security, and could be targeted by selecting heterogeneous formations at optimal injection rates. In this work, we experimentally quantify capillary heterogeneity trapping in a sandstone rock by performing core-flood experiments at different injection rates, mimicking the conditions in a field-project. We develop an analytical model which describes the proportion of capillary heterogeneity trapping in the system based on the rock properties, and find it agrees well with our experimental data. Our model can be used in the field to estimate trapping lengths and proportions, which would aid screening projects and operational design. 

To be an effective carbon removal technology, geological carbon storage requires injected carbon dioxide (CO$_2$) to remain securely trapped in the subsurface. Many mechanisms trap CO$_2$ in the subsurface, depending on the underlying geology and the relative forces present. Of interest here are the inherent variations in rock type and structure resulting in changes in CO$_2$ migration and accumulation. These subsurface variations lead to increased CO$_2$ trapping (capillary heterogeneity trapping) compared to a uniform site, generally enhancing storage security. However, as shown in this work, the amount of capillary heterogeneity trapping depends on the rate. In this work, we experimentally quantify capillary heterogeneity trapping in a sandstone rock by performing core-flood experiments at different injection rates, mimicking the conditions in a field-project. We develop an analytical model which describes the capillary heterogeneity trapping in the system based on the rock properties, and find it agrees well with our experimental data. Our model can be used in the field to estimate the trapping length and proportion of capillary heterogeneity trapping, which would aid screening projects and operational design.

\section{Introduction}

Geological heterogeneity strongly impacts flow and trapping during CO\textsubscript{2} sequestration \cite{Regnier2019, Kortekaas1985}. Recent advances in experimental core analysis show that both fluid migration and capillary trapping are sensitive to spatial heterogeneity on length scales (mm - cm) that are usually not resolved in field-scale flow simulations \cite{Jackson2018,Krevor2011, Ni2019}. Understanding the impact of this small-scale heterogeneity, at industrially relevant scales, is necessary to accurately predict CO\textsubscript{2} saturation distributions and security of CO$_2$ storage projects. 

Trapping of CO\textsubscript{2} by capillary forces is an important trapping mechanism during geological carbon storage. It occurs when water imbibes into rock partially saturated by CO\textsubscript{2} following injection. Within the pores of rocks ($\mu$m - mm), capillary forces snap-off and immobilise isolated ganglia of CO\textsubscript{2}. For water wet rocks, the amount of residually trapped CO\textsubscript{2} on the continuum scale following imbibition depends upon its initial saturation at the beginning of imbibition \cite{Krevor2015}. The Land model is a continuum representation of trapping frequently used to parameterise this relationship: 
\begin{equation}
    S_{nw,res} = \frac{S_{nw,init}}{1+CS_{nw,init}},
\label{eqn:Land}
\end{equation}
where the index ($nw$) refers to the non-wetting phase and $C$ is an empirical trapping constant \cite{Jackson2020REV,Krevor2020Chp}. The trapping constant is typically evaluated from imbibition core floods (e.g. \cite{Pentland2011}, \cite{Iglauer2011}).

Capillary heterogeneity trapping is an additional trapping mechanism resulting from heterogeneity in the continuum capillary pressure on the mm to m scale. These changes in capillary pressure result in local barriers to flow within the rock structure. They cause mm and cm scale increases in non-wetting phase saturation upstream of capillary barriers, accumulating as a continuous phase which remains effectively immobilised. This behavior can result in deviations from residual trapping models, such as the Land model \cite{Jackson2020REV}, \cite{Harris2021}. We note that capillary heterogeneity trapping has been called a variety of different terms, including local capillary trapping and capillary pinning \cite{Saadatpoor2010, zhang2025}. We use the term capillary heterogeneity trapping here, since it encapsulates the mechanism by which the trapping manifests; heterogeneity in capillary pressure characteristics.

Capillary heterogeneity has the potential to increase the security and capacity of CO\textsubscript{2} storage in underground aquifers \cite{Krevor2011, Harris2021}, however the evolution of this trapping with flow rate is poorly understood \cite{Reynolds2018, Niu2015}. Continuum theory predicts a decrease trapping with increasing flow rate due to the relative increase in viscous force over capillary forces \cite{Debbabi2017, Ringrose1993}. In part because of this sensitivity, capillary heterogeneity could explain deviations in measured initial-residual curves observed in the literature \cite{Dance2016, Bachu2013, Al-Menhali2016, Krevor2012} and scatter in the trapping data that is often attributed to experimental error \cite{Krevor2015, Reynolds2018, Ni2019}. 

The aim of this paper is to investigate the combined effects of continuum-scale capillary heterogeneity and flow rate on core flood evaluations of trapping. A series of drainage and imbibition core-flood experiments at three different flow rates, imaged using medical X-ray CT, demonstrate the rate-dependency of measured initial-residual relationships. An analytical model is derived to analyse these behaviours, highlighting the key parameters to include when modeling and upscaling capillary heterogeneity trapping.

\section{Analytical Model of Capillary Heterogeneity Trapping}

The 1D semi-analytical model presented below is developed from the conventional continuum theory of multiphase flow in porous media to describe the capillary heterogeneity trapped saturation distribution during steady state imbibition through a layered system. It is similar to the models presented by Duijn et al. (1995) and Dale and Ekrann (1997) that describe the saturation profile within 1D heterogeneous oil reservoirs \cite{Dale1997, Duijn1995} but extends them to the trapping stage of a CO\textsubscript{2} storage project where capillary, viscous and gravity effects may be important. The model is used to parameterise saturation discontinuities observed upstream of capillary pressure barriers, ensuring the capillary pressure continuity condition is fulfilled \cite{Debbabi2017}. We present an overview of the derivation here, referring the reader to the supporting information for further details. 

\begin{figure}
\centering
{\includegraphics[width=1\textwidth]{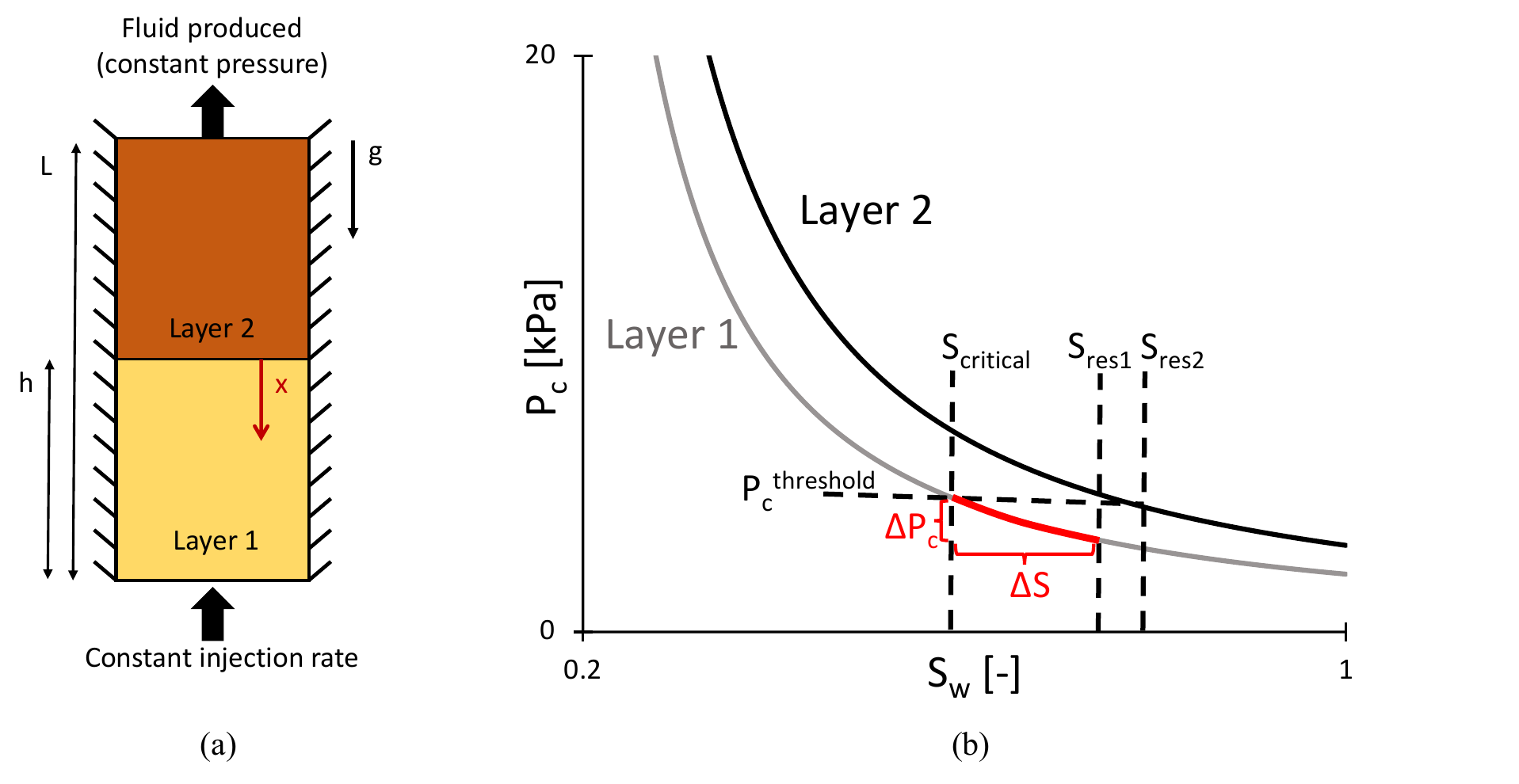}}
\caption{(a) Diagram of the model system with length \textit{L}. A region of high capillary entry pressure (layer 2) lies above a region of low capillary entry pressure (layer 1), with heterogeneity length \textit{h}. (b) The Brooks-Corey capillary pressure - saturation functions for the two layers of different entry pressure are shown. During the steady state imbibition of water, the critical CO\textsubscript{2} saturation (\textit{S\textsubscript{critical}}) forms in the low capillary entry pressure layer, upstream of a capillary pressure barrier, as a consequence of capillary pressure continuity.  \label{fig:Chp2_system}}
\end{figure}

The model considers a 1D medium formed of 2 homogeneous layers with flow perpendicular to those layers (Figure \ref{fig:Chp2_system}). Each layer has different properties, with the layer closest to the injection point having the lower capillary pressure. It is representative of an element in a composite core and vertical flow through two layers in a stratified aquifer \cite{Dale1997}. 

Carbon dioxide already exists in the system at the start of imbibition, unlike in the primary drainage case. During imbibition, water displaces the pre-existing CO{\textsubscript{2}}. Some of this CO{\textsubscript{2}} becomes trapped upstream of the heterogeneity interface due to the contrast in capillary pressure between the layers. Eventually, at steady state, only water is flowing. At this time the CO\textsubscript{2} in the upstream layer is immobilized. This is capillary heterogeneity trapping.  

Using fractional flow theory, the saturation gradient in a 1D homogeneous porous medium at steady state when only water is flowing, is given by:
\begin{equation}
    \frac{dS_w}{dx} = \frac{\frac{q_T\mu_w}{k_{rw}K} + \Delta\rho g}{P_c'}
    \label{eqn:satgrad}
\end{equation}
where $q_T$ is the total wetting phase Darcy flow velocity, \textit{K} is the absolute permeability, $k_{rw}$ is the wetting-phase relative permeability, $\mu_w$ is the wetting-phase viscosity, \textit{$\Delta\rho$} is the density difference between the fluids, \textit{g} is the component of gravitational acceleration in the direction of flow, and $P_c'$ is the gradient in capillary pressure with respect to wetting-phase saturation $(\frac{dP_c}{dS_w})$. The capillary pressure is defined as, \textit{P\textsubscript{c}(S\textsubscript{w}) = P\textsubscript{CO2} - P\textsubscript{w}}. 

We can obtain an equation describing the distribution of CO\textsubscript{2} trapped beneath the interface between the layers by integrating Equation \ref{eqn:satgrad}. This gives an expression for \textit{x}, the distance of the saturation from the heterogeneity, as a function of $S_w$ in the layer closest to the injection point.  
\begin{equation}
    x = h - \int_{S_w}^{S_{w,critical}}\frac{P_{c1}'}{\frac{q_T\mu_w}{k_{rw}K_1} + \Delta\rho g}ds_w
    \label{eqn:finalg}
\end{equation}
where $S_{nw,critical}$ = 1 - $S_{w,critical}$, is the critical CO\textsubscript{2} saturation immobilized at the interface between layer 1 and layer 2, following water imbibition. The saturation spatial distribution results in a capillary pressure gradient which ensures flux continuity. The extent of the trapped saturation behind the capillary barrier depends on the functions of the upstream region, in which the trapping takes place.

The critical CO\textsubscript{2} saturation trapped at the interface ($S_{nw,critical}$) is determined by applying the capillary pressure continuity boundary condition (see previous work by \cite{Dale1997, Duijn1995, VanDuijnDeNeefM.1996,  VanDuijn2016}). In order to do this we need a functional form for the imbibition capillary pressure in each layer. There is a limited literature on functional forms for imbibition capillary pressure so, for simplicity, we  use the Brooks-Corey model (originally derived for drainage) to parameterize the imbibition capillary pressure relationship, $P_c = P_e S^*_w{}^{-1/\lambda}$ where \textit{P\textsubscript{e}} is the capillary entry pressure and \textit{$\lambda$} is the Brooks-Corey parameter. \textit{$\lambda$} depends upon pore-size distribution and is a measure of pore-scale heterogeneity within a system \cite{Ni2019,BC}. The variable \textit{S\textsubscript{w}*} is the water saturation normalised with respect to the irreducible water saturation \textit{S\textsubscript{wirr}}, $S_w^* = \frac{S_w-S_{wirr}}{1-S_{wirr}}$. Experimental data suggest that the shape of imbibition capillary pressure curves is not significantly different from drainage curves \cite{Kleppe1997}, \cite{Wang2016}, although the analysis is valid for other functional forms. 

Using this we can calculate the capillary pressure at the interface between layers 1 and 2, 
\begin{equation}
    P_c^{threshold} = P_{e2}(S^{*}_{w2,x=h})^{-1/\lambda}
    \label{eqn:pct1}
\end{equation}
\begin{equation}
    P_c^{threshold} = P_{e1}(S^{*}_{w1, critical})^{-1/\lambda} 
    \label{eqn:pct2}
\end{equation}

Hence the critical wetting phase saturation at that interface is calculated
\begin{equation}
    S^{*}_{w1, critical} = (\frac{P_{e2}}{P_{e1}})^{-\lambda}S^{*}_{w2,x=h}
    \label{eqn:scrit}
\end{equation}
%S^{*}_{w1,x=H}
The maximum capillary heterogeneity trapped saturation thus depends on the ratio between the upstream and downstream capillary entry pressures. The CO{\textsubscript{2}} saturation in the upper layer (layer 2) is assumed to be displaced to its residual value on imbibition in the simple two-layer system ($S^{*}_{w2,x=h} =S^{*}_{w,res2}$).

From Equation \ref{eqn:finalg} we see that the extent of the trapped saturation behind the capillary barrier also depends on the wetting phase relative permeability function ($k_{rw}$). In this work we assume universal relative permeability curves over the domain, a common modelling assumption supported by Burdine's theory \cite{Li2015}. They are parameterised using the Corey power law model, $k_{rw} = k_{rw}^eS^*_w{}^n$, where \textit{n} is the Corey parameter and \textit{k\textsubscript{rw}\textsuperscript{e}} is the end point wetting phase relative permeability \cite{Niu2015}. Substituting this functional form, along with the Brook's Corey model for capillary pressure, into Equation \ref{eqn:final} gives,
\begin{equation}
    x = h - \int_{S_w}^{S_{w1,critical}}\frac{\frac{P_{e1}}{\lambda} s^*_w{}^{-1/\lambda -1}}{\frac{q_T\mu_w}{K_1k_{rw}^es^*_w{}^n} + \Delta\rho g}d{s_w}
    \label{eqn:finalS}
\end{equation}
Equation \ref{eqn:finalS} is solved numerically, although it can be solved analytically if simple linear functions for capillary pressure and relative permeability are chosen (that is, if \textit{n} and \textit{$\lambda$} are 1). 

\subsection{Dimensionless trapping length-scale}
\label{secxH}

Equation \ref{eqn:finalS} provides the system parameters which influence capillary heterogeneity trapping and so can be used to evaluate under which conditions laboratory measurements of initial-residual trapping relationships may be affected by capillary heterogeneity within the core. Reducing these parameters to dimensionless numbers provides insights into the relative importance of gravity, capillary and viscous effects on the core flood and hence on the initial-residual trapping measurements. This indicates at which storage sites and conditions capillary heterogeneity trapping is expected to be significant.

Traditional dimensionless numbers often fail to describe systems displaying capillary heterogeneity trapping due to the exclusion of key parameters incorporating heterogeneity \cite{Ni2021}. Using the analytical model, we define a novel dimensionless number, the dimensionless trapping length scale,
\begin{equation}
    \overline{x_T} = \frac{x_T}{h} = \frac{h-x_{S=S_{res1}}}{h} 
    \label{eqn:xT}
\end{equation}
where \textit{h} is the thickness of the high permeability layer in which the trapping takes place (layer 1), \textit{x\textsubscript{T}} is the distance over which capillary heterogeneity trapping occurs, and $x_{S=Sres1}$ is the location at which the capillary heterogeneity trapped gas saturation becomes zero (i.e. the saturation equals the residual saturation).  

Defining the viscous to capillary number $N_{v/c}$ as
\begin{equation}
    N_{v/c}=\frac{\lambda q_T\mu_w}{K_1 k_{rw}^e P_{e1}}
    \label{eqn:Nvg}
\end{equation}
and the macroscopic gravity to capillary number $N_{g/c}$ as
\begin{equation}
    N_{g/c}=\frac{\lambda\Delta\rho g}{P_{e1}}
    \label{eqn:NPc}
\end{equation}
Substituting for these into Equation {\ref{eqn:finalS}} gives  
\begin{equation}
    \overline{x_T} = \frac{1}{h} \int_{S_{res1}}^{S_{w1,critical}}\frac{s^*_w{}^{-1/\lambda-1}} {\frac{N_{v/c}}{s^*_w{}^n} + N_{g/c}} ds_w.
    \label{eqn:final}
\end{equation}
The dimensionless trapping length incorporates key parameters such as heterogeneity contrast and capillary pressure characteristics. This equation shows that the larger the density difference then the larger $N_{g/c}$ and the smaller the dimensionless trapping length. If the flow is horizontal then the trapping length depends primarily on the viscous to capillary number: the larger the flow rate then the smaller the trapping length.

\section{Experimental}
We investigated the impact of capillary heterogeneity on initial-residual trapping in a real system by performing a series of core floods in an heterogeneous Bentheimer core. A medical X-ray CT scanner was used to determine the porosity distribution and capillary heterogeneity trapped gas saturation along the core. Additional experimental details and figures are provided in the supporting information. 

\subsection{Materials}

\begin{figure}[h!]
\centering
{\includegraphics[width=0.7\textwidth]{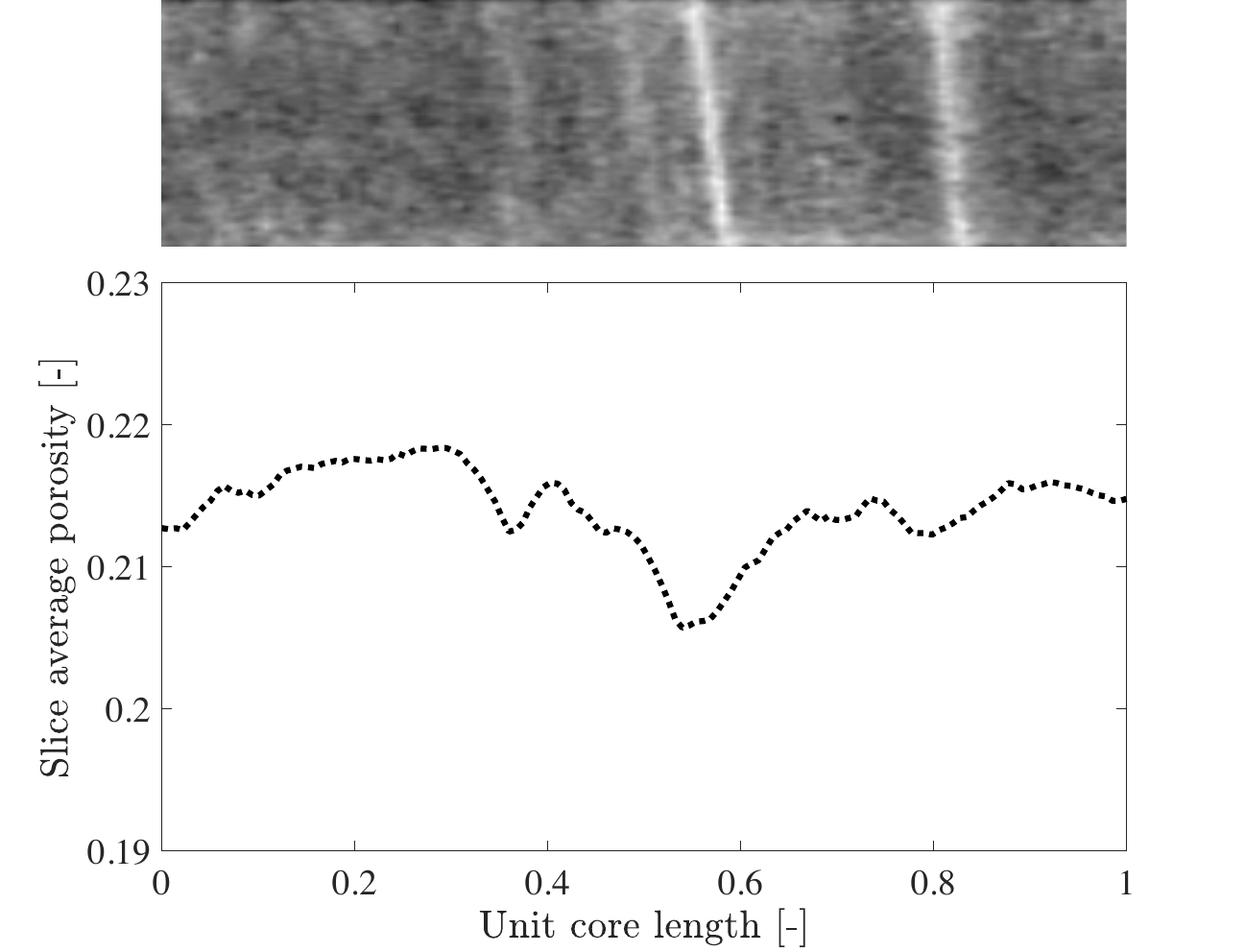}}
\caption{Grey scale visualisation of CT number for a central YZ slice (top) and slice average porosity along the normalised core length (bottom), calculated from medical-CT scans of the Bentheimer core. \label{fig:Chp4_A3}}
\end{figure}

The Bentheimer sandstone core was chosen as it had layers perpendicular to the axis of the core that were clearly visible to the naked eye. The average core porosity was $0.214 \pm 0.003$,  determined from the CT numbers using the method described in Krevor et al. (2012) \cite{Krevor2012}. Variations in porosity along core length are shown in Figure \ref{fig:Chp4_A3}. In the greyscale image, bright regions highlight the major heterogeneities, probably due to iron oxide deposits. The average absolute permeability ($K$) was estimated to be 0.94 D \cite{Reynoldsthesis2016}.

N$_2$ and DI water were used as working fluids in analogy to a supercritical CO$_2$-brine system \cite{Niu2015}. The water viscosity, nitrogen viscosity and interfacial tension were estimated respectively as $\mu_w$ = $1.06 \times 10^{-3}$ Pa$\cdot$s, $\mu_{N_2}$ = $1.96 \times 10^{-5}$ Pa$\cdot$s and $\gamma_{N_2-w}$ = $0.067$ N/m for the experimental conditions\cite{Jackson2020}. 

\subsection{Method} 

N$_2$-DI water core-floods were performed to measure the initial-residual saturation relationship at elevated pressure (10 MPa pore-pressure) and ambient temperature. First a drainage core-flood was performed where 100\% N$_2$ was injected. The fluid distribution was then imaged once the system reached steady state. Due to the capillary end effect and the impact of heterogeneity, a range of initial non-wetting phase saturations were observed within the core \cite{Niu2015}. An imbibition core-flood with 100\% water injection was then carried out to obtain the residual saturation. By correlating the final saturation in each voxel to the saturation prior to imbibition, the initial-residual saturation relationship was obtained \cite{Krevor2015}. 

This experimental sequence was performed at three imbibition flow rates with a constant drainage flow rate between each run, Table \ref{tab:rate_exp}, to isolate the impact of imbibition rate on capillary trapping. The corresponding pore-scale capillary number, $N_{c} = \frac{q\mu}{\gamma}$, where $q$ Darcy velocity and $\mu$ viscosity of the invading phase was also calculated for each experiment. Full details of the experimental methodology are provided in the supporting information.  

\begin{table}[ht]
\begin{center}
    \begin{tabular}{ |c|c|c|c|c| } 
        \hline
        \rule{0pt}{3ex} 
         & \multicolumn{2}{|c|}{Rate [cm$^3$min$^{-1}$]} & \multicolumn{2}{|c|}{Capillary number [-]} \\
         \cline{2-5}
         \rule{0pt}{3ex} 
        Run & Drainage & Imbibition & Drainage & Imbibition \\
         \hline
         \rule{0pt}{3ex} 
        1 & 10 & 0.05 & 4.1$\times$10\textsuperscript{-8} & 1.$\times$10\textsuperscript{-8} \\
        2 & 10 & 0.1 & 4.1$\times$10\textsuperscript{-8} & 2.2$\times$10\textsuperscript{-8} \\
        3 & 10 & 0.5 & 4.1$\times$10\textsuperscript{-8} & 1.1$\times$10\textsuperscript{-7} \\
        \hline
    \end{tabular}
     \caption{\label{tab:rate_exp} Injection rates used in the initial-residual trapping experiments and associated experimental pore-scale capillary number. }
\end{center}
\end{table}
The saturation distributions were measured using medical CT scanning with voxel size 0.09 mm (x,y) and 1 mm (z - slice thickness). In order to reduce uncertainties in voxel-scale saturations, the voxels were coarsened by averaging over 20 voxels in the x-y directions. The resulting coarsened voxels had dimensions of 2 $\times$ 2 $\times$ 1 mm. These values were chosen to give the coarsened voxels well-defined Darcy-scale flow properties, with representative elementary volume (REV) similar to those in the literature \cite{Jackson2020REV,Herring2013,Pini2013}. The voxel-scale N\textsubscript{2} saturation values reported in this study are from these coarsened voxels. 

Jackson et al. (2020) also carried out initial-residual trapping core-flood experiments on a heterogeneous Bentheimer sample similar to the core used in these experiments \cite{Jackson2020REV}. Their cores had similar perpendicular layered heterogeneity as they came from the same slab. Therefore, the results of Jackson et al. (2020) are used as a literature comparison to this study. 

\subsection{Quantifying Capillary Heterogeneity Trapping}

The 1D analytical model outlined was applied to the results of the core-flood experiment to test its utility in real 3D systems. Equation \ref{eqn:finalS} was calculated with the gravity term excluded as the core-flood was performed in the horizontal orientation. The input parameters for the model, detailed in Table \ref{tab:BenAM}, were taken from the experiment and from Jackson et al. (2020) \cite{Jackson2020REV}. 

\begin{table}[ht]
\begin{center}
    \begin{tabular}{ |c|c|c| } 
        \hline
        \rule{0pt}{3ex} 
        Parameter & Value & Unit \\
         \hline
         \rule{0pt}{3ex} 
        $K$ & 9.87x10$^{-13}$ & m$^2$\\
        $k_{rw}^e$ & 1 & -  \\
        $n$ & 4 & -  \\
        $\mu_w$ & 0.001 & Pa$\cdot$s \\
        $\lambda$ & 2.7 & - \\
        $S_{wirr}$ & 0.08 & - \\
        $C$ & 1.24 & - \\
        $h$ & 0.07 & m \\
        $q$ & 7$\times$10$^{-7}$ - 7$\times$10$^{-6}$ & ms$^{-1}$ \\ 
        $P_e$ & 1.2 - 3.7 & kPa \\
        $\Delta P_{e,h}$ & 0.5 - 1.5 & kPa \\
        \hline
    \end{tabular}
     \caption{\label{tab:BenAM} Parameters used in the analytical model, with values obtained from the experiment and from Jackson et al. (2020) \cite{Jackson2020REV}. $P_e$ is the capillary entry pressure of the layers and $\Delta P_{e,h}$ is the change in capillary entry pressure over the heterogeneity. }
\end{center}
\end{table}

The capillary entry pressure of the different layers in the core was not directly measured. Instead, based off Brooks-Corey predictions from Jackson et al. (2020), the capillary entry pressure is estimated as 3.7 kPa and 1.2 kPa for drainage and imbibition respectively. In addition, the change in the capillary pressure over the heterogeneity was estimated as 1.5 kPa \cite{Jackson2020REV}. Due to the uncertainty in assigning capillary entry pressure to the different layers within the core, analytical solutions were calculated for a range of values.  

To understand the factors influencing capillary heterogeneity trapping, the total trapped saturation was compared to that trapped by pore-scale residual saturation alone, with differences attributed to capillary heterogeneity trapping. The pore-scale residual trapping was estimated from Equation \ref{eqn:Land} using $C_{max}$, the maximum Land trapping coefficient obtained experimentally. The proportion of trapping resulting from capillary heterogeneities was then quantified as,
\begin{equation}
     H\% = \frac{\textrm{System average capillary heterogeneity trapped saturation}}{\textrm{System average total trapped saturation}} \times 100.
        \label{eqn:percent}
\end{equation} 
\section{Results}

\subsection{Initial-residual trapping relationship}

\begin{figure}[h!]
\centering
{\includegraphics[width=1.0\textwidth]{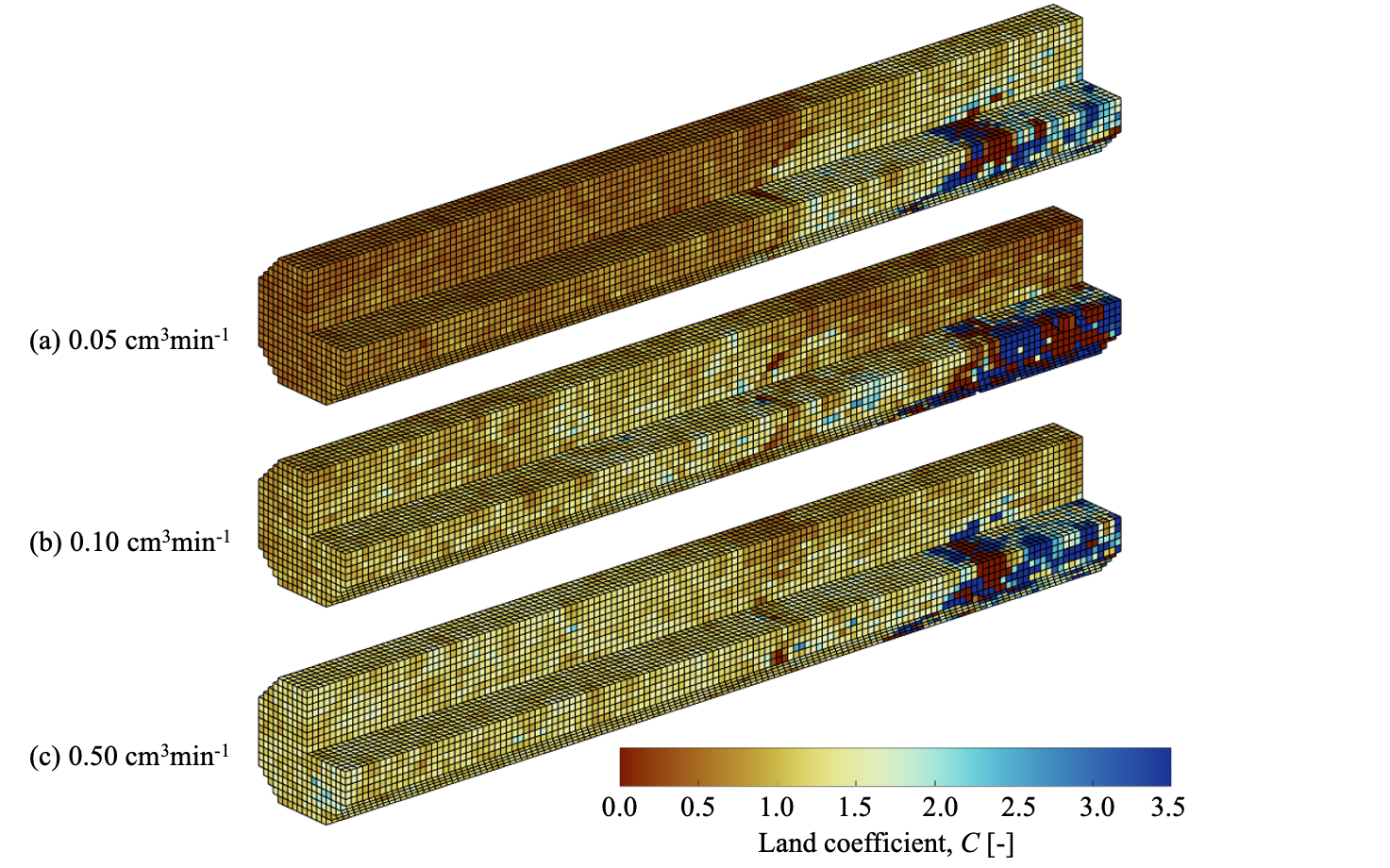}}
\caption{Voxel-scale Land trapping coefficient within the Bentheimer sample for three flow rate experiments, displayed in 3D. Flow is from left to right in all images. \label{fig:Chp4_VoxelLandBen_3D}}
\end{figure}

Examination of Figure \ref{fig:Chp4_A3} shows that there is a zone of reduced porosity approximately 0.55 of the distance between inlet and outlet. Capillary heterogeneity trapping is observed experimentally within the heterogeneous Bentheimer core, with an increase in non-wetting phase saturation upstream of this location. Figure \ref{fig:Chp4_VoxelLandBen_3D} plots the voxel-scale Land trapping coefficient in 3D for the three different flow rate experiments. A clear rate dependency in the trapping relationship is observed. These results show that as flow rate is decreased, the Land trapping coefficient decreases upstream of the heterogeneity as a result of capillary heterogeneity trapping. The capillary heterogeneity trapped saturation remains stable for many pore volumes of brine injected, establishing itself as a long term trapping mechanism rather than a transient effect. 

\begin{figure} 
\centering
{\includegraphics[width=0.65\textwidth]{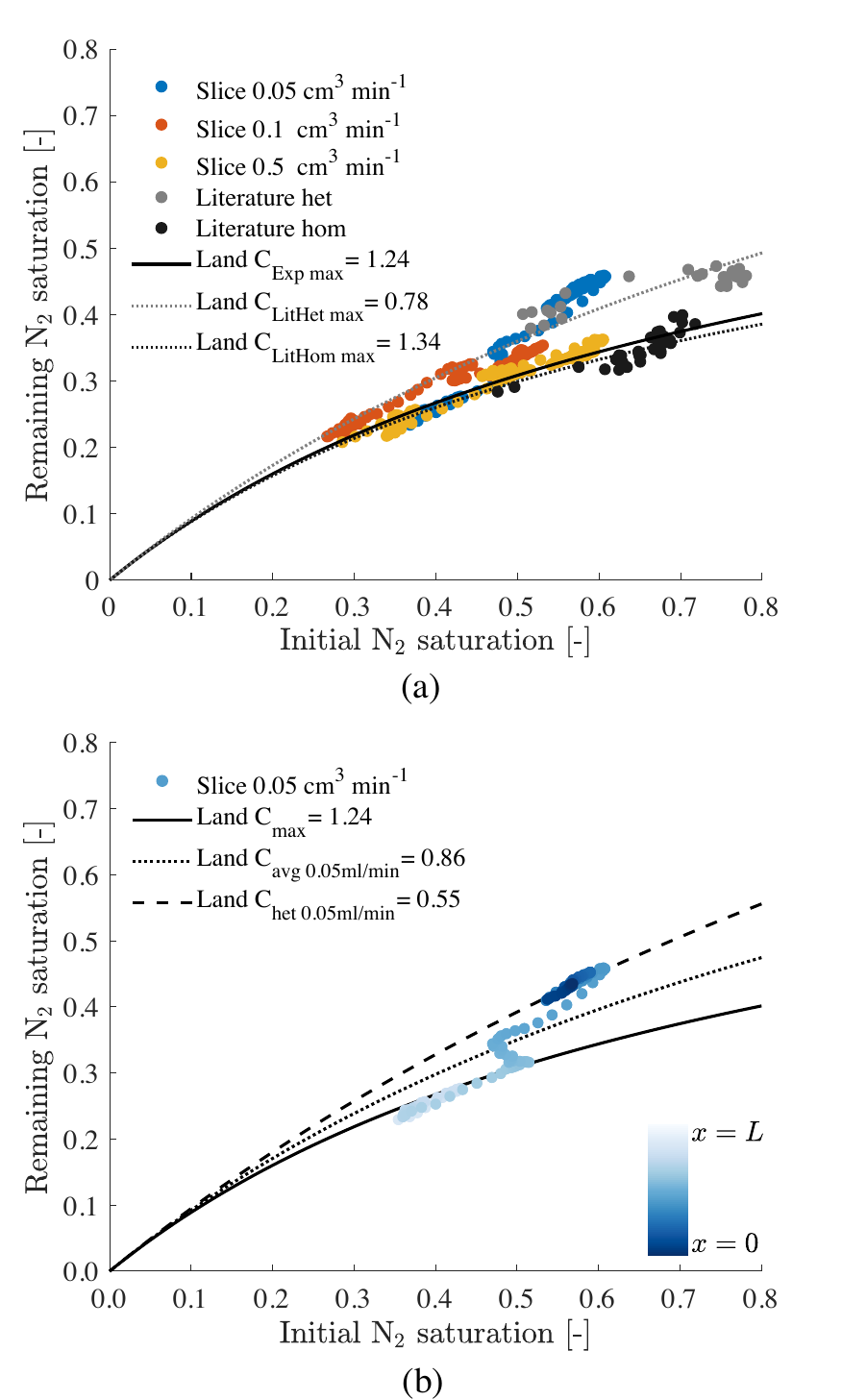}}
\caption{(a) Slice average initial-residual N\textsubscript{2} saturation for three flow rate experiments in the Bentheimer sample (blue, orange and yellow markers). The maximum Land trapping relationship, calculated as the slice average relationship from the high flow rate (0.5 cm$^3$min$^{-1}$) experiment, is plotted as a solid line ($C = 1.24 \pm 0.13$). In addition literature comparisons from Jackson et al. (2020) for a homogeneous and heterogeneous Bentheimer \cite{Jackson2020REV} are shown (grey and black markers). The average trapping relationships for the homogeneous $C=1.34$ and heterogeneous $C=0.78$ literature samples are also shown (dotted lines). (b) Slice average initial-residual N\textsubscript{2} saturation for the imbibition rate 0.05 cm$^3$min$^{-1}$ experiment. The colour gradient shows where the saturations occur along the core, from inlet ($x$=0) to outlet ($x$=$L$). The Land trapping relationship upstream of the heterogeneity $C=0.55$ (dashed line), as well as the maximum $C=1.24$ (solid line) and average $C=0.86$ (dotted line) Land trapping relationships are displayed  \label{fig:Chp4_expCHT}}
\end{figure}

The experimentally measured initial-residual trapping relationships are a function of water imbibition rate, as expected from the analytical model. Figure \ref{fig:Chp4_expCHT}a plots the initial-residual N$_2$ saturation for three different imbibition flow rates, with the Land trapping relationship defined according to Equation \ref{eqn:Land}. The maximum Land trapping relationship, $C = 1.24 \pm 0.13$ with one standard deviation uncertainty, best characterises the primarily pore-scale phenomenon of residual trapping \cite{Reynolds2018}. This is calculated from the high flow rate (0.5 cm$^3$min$^{-1}$) experiment, which obeys a Land trapping relationship with minimal scatter, implying the results are a good descriptor of the pore-scale residual trapping relationship. Further plots of the voxel level initial-residual relationship are given in the supporting information. 

Deviations in the residual saturation above the maximum Land trapping relationship are attributed to capillary heterogeneity trapping. Figure \ref{fig:Chp4_expCHT}a shows that as the imbibition flow rate is decreased, there is a bigger difference between the remaining saturation and the maximum Land trapping relationship, due to the increase in trapping by capillary heterogeneities. 

The experimental results were compared to the trapping experiments carried out on a homogeneous and heterogeneous Bentheimer sample in Jackson et al. (2020) \cite{Jackson2020REV}. Figure \ref{fig:Chp4_expCHT}a shows the high flow rate (0.5 cm$^3$min$^{-1}$) experiment displays a similar trapping behaviour to the Jackson et al. (2020) homogeneous sample. The Land trapping relationship from the high flow rate experiment is $C=1.24 \pm 0.13$, whilst the homogeneous experiment from Jackson et al. (2020) has a best fit Land correlation (slice average) of $C = 1.34 \pm 0.12$, with both curves falling within the 1 standard deviation of the other. In contrast the low flow rate (0.05 cm$^3$min$^{-1}$) experiment displays a similar trapping behaviour to the Jackson et al. (2020) heterogeneous experiment. Jackson et al. (2020) used a lower Land trapping coefficient of $C=0.5$ to parameterise the trapping in this sample. This trapping relationship is comparable to our low flow rate experiment (0.05 cm$^3$min$^{-1}$) fitted upstream of the heterogeneity, $C=0.55 \pm 0.05$.

Comparing the results of these experiments to the analogous homogeneous and heterogeneous literature samples highlights the variable impact of heterogeneity on trapping. The changing nature of capillary heterogeneity trapping in our experiment suggests that, depending on the flow conditions used, the initial-residual trapping relationship may behave more similarly to a heterogeneous or homogeneous sample. This highlights the difficulties in measuring trapping relationships in the laboratory and then applying them to field-scale projects where imbibition flow rates may be different. 

At low flow rates, the contrast between the maximum and minimum Land trapping coefficients determined from different sections of a heterogeneous core is much higher. Figure \ref{fig:Chp4_expCHT}b highlights the difference between extracting the maximum Land trapping coefficient ($C=1.24 \pm 0.13$) and average Land trapping coefficient ($C=0.86 \pm 0.37$) from experimental data at low flow rate (0.05 cm$^3$min$^{-1}$). A clear variation in the trapping relationship is observed for initial saturations greater than approximately $S_{N2}=0.5$, corresponding to regions upstream of the heterogeneity. The average Land trapping coefficient for the region upstream of the heterogeneity is $C=0.55 \pm 0.05$, with the higher remaining saturation indicative of capillary heterogeneity trapping. Isolating this region, the small uncertainty in the trapping relationship (9\%), suggests this data region may be used to capture the capillary heterogeneity trapping relationship. Examining the data nearest the core inlet in Figure \ref{fig:Chp4_expCHT}b suggests a linear trapping relationship may be better suited to parameterise this region \cite{Krevor2011}.

Based on these results, we suggest that a high standard deviation in the Land trapping coefficient determined from low flow rate experiments may be indicative of capillary heterogeneity trapping within the core. For example in our results, one standard deviation in the average Land trapping coefficient is 44\% for low flow rate experiment (0.05 cm$^3$min$^{-1}$), compared to 10\% for the high flow rate experiment (0.5 cm$^3$min$^{-1}$). This suggests that the degree of uncertainty in the average trapping relationship may be used as a measure of the potential presence of capillary heterogeneity trapping in a system. 

%%%%%%%%%%%%%%%%%%%%%%

\subsection{Rate dependent capillary heterogeneity trapping}

\begin{figure}
\centering
{\includegraphics[width=0.7\textwidth]{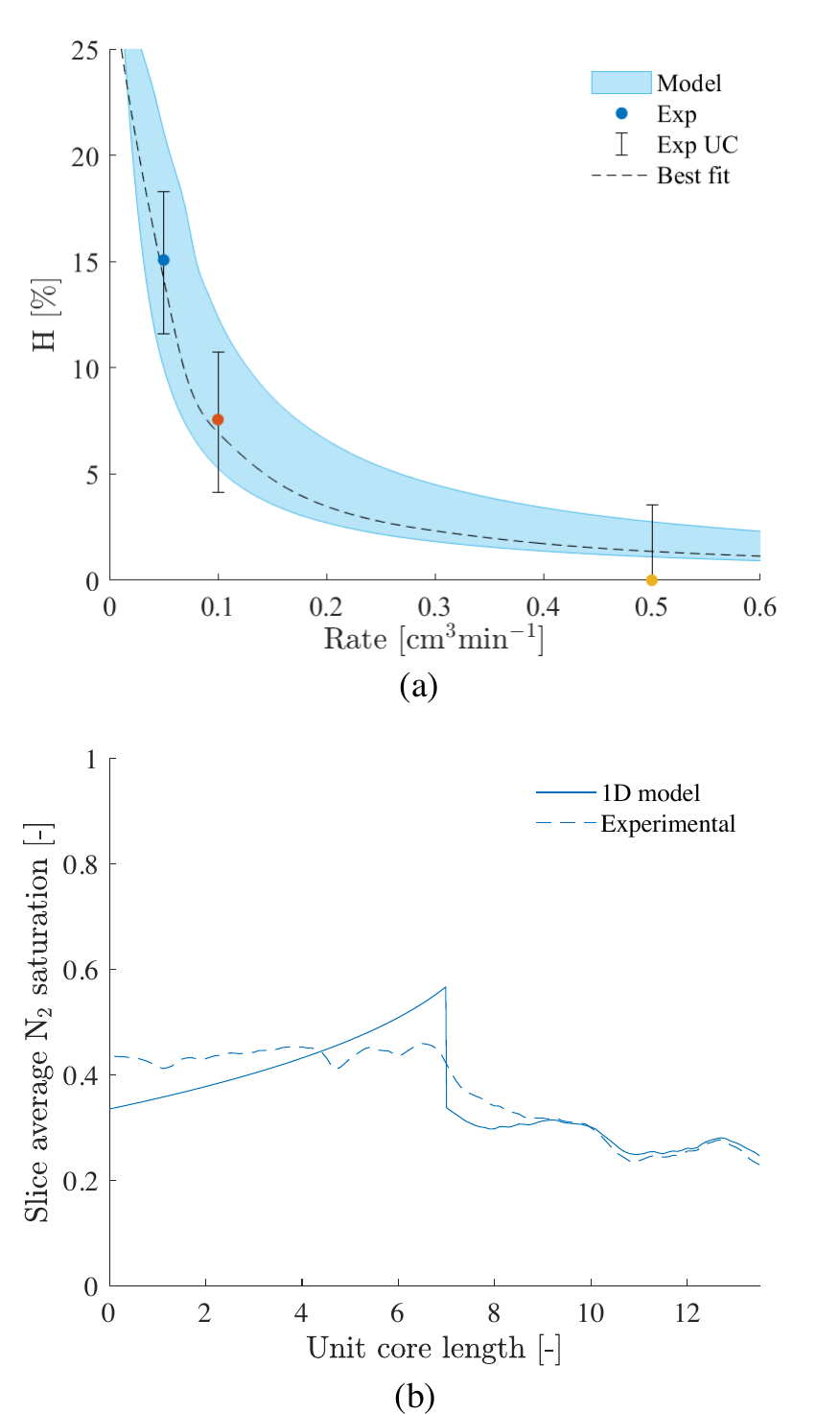}}
\caption{(a) Proportion of capillary heterogeneity trapping ($H$\%) over different flow rates, estimated for the Bentheimer experiments from the analytical model for a range of capillary entry pressures. In addition, the experimental results (The proportion of capillary heterogeneity trapping in the system as a function of water imbibition rate, with $\pm$ 1 standard deviation of uncertainty in $C$ shown.) and associated uncertainties are shown. The capillary pressure parameters in the analytical model are optimised to provide a good fit to the experimental data. (b) Analytical solution of trapped saturation compared to experimental results, for flow rate 0.05 cm$^3$min$^{-1}$. The capillary pressure parameters in the analytical model are optimised to provide a good fit to the proportion of capillary heterogeneity trapping from the experimental data. 
\label{fig:Chp4_AM}}
\end{figure}

The capillary heterogeneity trapped saturation is quantified by comparing the remaining saturation trapped to that trapped from pore-scale residual trapping alone. Figure \ref{fig:Chp4_AM}a shows the proportion of capillary heterogeneity trapping $(H)$ within the experiments as a function of water imbibition rate. A significant uncertainty in the proportion of capillary heterogeneity trapping within the system results from uncertainty in the maximum Land trapping coefficient extracted. The uncertainty in fitting the maximum initial-residual saturation relationship (10\%) is much larger than the slice-average saturation uncertainty due to experimental precision ($< 0.15\%$) \cite{Pini2012, Jackson2018}.

A good agreement is observed between the proportion of capillary heterogeneity trapping estimated from the analytical model and obtained from the experiments as a function of rate, Figure \ref{fig:Chp4_AM}a. The analytical model was run over the range of rates and capillary pressures outlined in Table \ref{tab:BenAM}, resulting in a range in the amount of capillary heterogeneity trapping. The initial saturation in the analytical model was assumed to be the core average initial saturation ($S_{init} = 0.5$). The results of the analytical model fall mainly within the uncertainty bounds in estimating the proportion of capillary heterogeneity trapping from the experimental results. This shows that the 1D model is able to estimate the proportion of capillary heterogeneity trapping within a real 3D core when the heterogeneity pattern is approximately 1D. The difference between the analytical model and experimental results is highest at high flow rates due to differences in how the pore scale residually trapped gas is handled at the boundary. Figure \ref{fig:Chp4_AM}a shows the best fit between the analytical model and experimental results. This was obtained by selecting a capillary entry pressure and capillary pressure contrast between the low and high capillary pressure layers that best matched the experimental results; $P_{e1}=$ 3 kPa, $\Delta P_{c}=$ 0.8 kPa.  

Figure \ref{fig:Chp4_AM}b overlays the slice average saturation predicted from the best fit 1D model with the experimental results at low flow rate (0.05 cm$^3$min$^{-1}$). In this instance, the slice average initial saturation from the experiment was used. It should be noted that whilst the 1D model provides a good match to the proportion of capillary heterogeneity trapping in the system, the model can only estimate the 1D saturation distribution because in reality the 3D core is more complex, with other small-scale heterogeneity present and a gradation in grain size at the heterogeneity. 

\begin{figure}[h!]
\centering
{\includegraphics[width=0.9\textwidth]{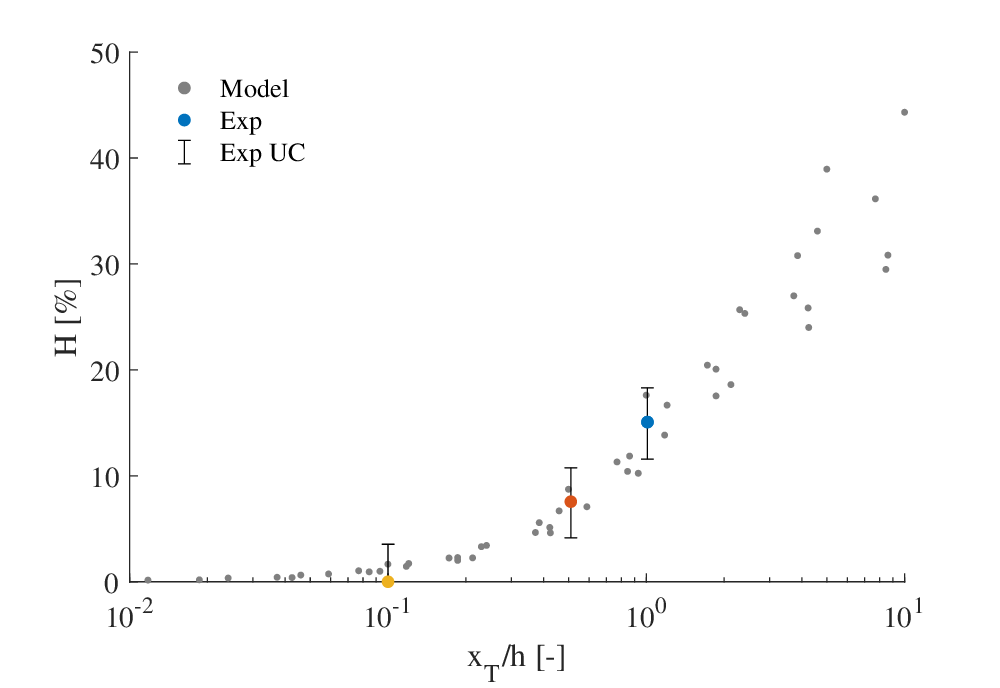}}
\caption{Proportion of capillary heterogeneity trapping ($H$\%) as a function of dimensionless trapping length ($x_T/h$). The analytical model is run over many orders of magnitude, including varying the Brooks-Corey ($\lambda$ \{0.5-1.5\}) and Corey parameters (\textit{n} \{4-7\}). The dimensionless trapping length is estimated for the Bentheimer experimental system with optimised capillary pressure, plotted against the proportion of capillary heterogeneity trapping calculated from the experimental results. Error bars indicate uncertainty in the proportion of capillary heterogeneity trapping calculated from experiments, resulting from uncertainty in defining the pore-scale residual trapping relationship. \label{fig:Chp4_xHBen}}
\end{figure}

Equation \ref{eqn:xT} defined the dimensionless trapping length as a means of predicting the proportion of capillary heterogeneity trapping in a system. The dimensionless trapping length was estimated for the Bentheimer system using the best fit capillary pressure described in the previous paragraph, Table \ref{tab:BenCHT}. Figure \ref{fig:Chp4_xHBen} shows the dimensionless trapping length can estimate the proportion of capillary heterogeneity trapping within the real rock core-flood at the experimental conditions used, as well as for different rates, Brooks-Corey parameter ($\lambda$) and Corey parameter ($n$). This suggests that the dimensionless trapping length is a good measure for estimating capillary heterogeneity trapping within a system.

\begin{table}[ht]
\begin{center}
    \begin{tabular}{ |c|c|c|c| } 
        \hline
        \rule{0pt}{3ex} 
        Rate [cm$^3$min$^{-1}$] & $x_T/h$ & $H$\%  predicted & $H$\% experiment \\
         \hline
         \rule{0pt}{3ex} 
        0.05 & 1.01 & 15 & 15\\
        0.1 & 0.51 & 7.6 & 7.6  \\
        0.5 & 0.10 & 1.1 & 0  \\ 
        \hline
    \end{tabular}
     \caption{\label{tab:BenCHT} Dimensionless trapping length calculated for the different experimental flow rates. The proportion of capillary heterogeneity trapping ($H$\%) is predicted using Figure \ref{fig:Chp4_xHBen} and is shown to agree well with the proportion calculated from the experimental results, within the uncertainty in characterising the pore-scale residual trapping relationship from experiments.}
\end{center}
\end{table}

\section{Discussion}

\subsection{Implications for core-flood characterisation}

%P1 - Care needed for exp in het systems
The experimental results show that capillary heterogeneity trapping depends upon the flow rate used in the core flood: the higher the flow rate then the lower the amount of capillary heterogeneity trapping. This is consistent with the 1D model for layered heterogeneous systems (Equation \ref{eqn:finalS}). In contrast, in homogeneous rocks, Niu et al. (2015) showed that the average Land trapping coefficient is rate independent \cite{Niu2015}. Thus, a key indicator that an experiment is influenced by capillary heterogeneity is that the initial-residual trapping curves change with flow rate. The 1D model (Equation \ref{eqn:final}) shows that, for horizontal corefloods, the ratio of capillary to viscous forces is key to determining the amount of capillary heterogeneity trapping. In addition, the critical saturation trapped at the heterogeneity boundary is controlled by the ratio of the capillary entry pressures of the layers.  

%P2 - average Land
The average trapping relationship is typically extracted from total initial-residual core-flood measurements \cite{K43, Krevor2012, Dance2016}. This core-scale trapping coefficient, in most cases, is an upscaled value that incorporates capillary heterogeneity trapping within the core. Jackson et al. (2019) have shown that a Land trapping model determined in this way may over predict the disconnected pore-scale residually trapped saturation in heterogeneous cores, whilst under-predicting the connected saturation caused by high entry pressure layers building-up the non-wetting phase upstream \cite{Jackson2020REV}. 

%P3 - Pore scale residual
Given that most rock cores probably contain small-scale heterogeneity, care is needed when undertaking core measurements of the initial-residual trapping curve. One approach is to perform experiments at a high capillary number to suppress capillary heterogeneity trapping. Alternatively, based on the findings of this work, the maximum Land trapping coefficient (rather than the average) may be extracted from experimental data to capture the impact of pore-scale residual trapping alone. Reynolds et al. (2018) proposed a similar approach of characterising a lower bound on the trapping, associating trapped saturations greater than the Land model to local capillary heterogeneities \cite{Reynolds2018}. Characterising capillary heterogeneity trapping is challenging due to the dependence on system conditions. Experiments should be carried out under characteristic conditions to quantify the impact of heterogeneity on trapping in the field. Alternatively, initial-residual trapping experiments carried out over different flow rates demonstrate the range in potential capillary heterogeneity trapping.

\subsection{Implications at the field-scale}

% P1 - Rate dependency at field-scale

The purpose of measuring initial-residual trapping relationships in the laboratory is to provide input data for field-scale simulations. Such simulations are used to design storage schemes to maximise storage security, including evaluating the amount of residually and capillary heterogeneity trapped CO\textsubscript{2}. In practice imbibition does not occur at a constant rate during field scale CO\textsubscript{2} storage as the imbibing water is driven both by capillary forces and buoyancy \cite{Mouche2010}. The relative magnitude of these varies across the field and with time, depending on the initial CO\textsubscript{2} saturations resulting from injection and the rate of CO\textsubscript{2} plume segregation from the water post injection. Using laboratory measurements of average initial-residual curves determined from low rate experiments may result in over-prediction of residually trapped CO\textsubscript{2} saturations. Conversely, using initial-residual curves determined at high flows will mean the simulation may underestimate the amount of core-scale capillary heterogeneity trapped CO\textsubscript{2}. 

A further consideration is that capillary heterogeneities occur in aquifers from all scales from the sub-core scale to greater than the metre scale \cite{Jackson2020}. These are not typically resolved in field scale simulation where the vertical dimensions of grid blocks may be 1-10m. These sub-grid block and greater than core scale heterogeneities will result in further capillary heterogeneity trapping. Upscaling is needed to convert laboratory measurements of initial-residual trapping data to the field. Given the rate dependency of capillary heterogeneity trapping we hypothesize that these upscaled functions may need to incorporate a rate dependent Land trapping parameter. This could be achieved by proving tables of the Land trapping coefficient for different rates or possibly devising a different formulation altogether.  

% P2 - Need to model at field-correctly, implications. 
Capillary heterogeneity trapping has important consequences in field-scale projects, particularly at sites without a laterally extensive caprock. In these composite confining systems, capillary entry pressure variations are used to baffle flow, controlling the overall footprint of the CO\textsubscript{2} plume \cite{Bump2023}. Therefore, identifying the mechanism by which saturation is trapped at the core-scale and extracting the appropriate trapping relationship when upscaling to the field is critical. For example, considering an initial saturation in the range $S_{CO2,init}$ 0.3–0.6, extracting the average or maximum Land trapping coefficient from the experimental results would correspond to a difference of 9–15\% in the capillary trapped saturation. In a site injecting 1 Mt/yr this would equate to 90–150 kt of CO\textsubscript{2} a year which could be incorrectly allocated depending on the Land trapping relationship extracted from core-scale experiments. Whether the CO\textsubscript{2} is trapped by pore-scale residual trapping or larger scale capillary heterogeneity trapping has important implications for site operators considering storage capacity and security of a site.

\section{Conclusion}

Capillary heterogeneity trapping has been directly observed in a heterogeneous Bentheimer core, with a local increase in non-wetting phase saturation upstream of the capillary pressure barrier. Experimentally, the proportion of capillary heterogeneity trapping is shown to be inversely proportional to brine imbibition rate. The observed initial-residual trapping relationships  behaved similarly to literature relationships for homogeneous or heterogeneous cores, depending on flow rate.  

The analytical model outlined can be used identify the key parameters influencing capillary heterogeneity trapping. The experimental data is observed to match with analytical model predictions. In future work, the analytical model could be used to estimate the proportion of capillary heterogeneity trapping within heterogeneous core-floods. The dimensionless trapping length is identified as the key dimensionless number to describe capillary heterogeneity trapping. In addition, by matching the analytical solution to measured initial-residual trapping characteristics, further details of the core may be obtained such as the capillary pressure of the layers or even the parameterisation of relative permeability. 

Further work is needed to develop a parameterisation of the initial-residual saturation relationship that can incorporate the influence of rate on sub-core scale capillary heterogeneity trapping and can also be used in upscaling to the field scale.

\section*{Open Research}
Data and software (analytical model, data analysis, data visualization, and model output) associated with this work may be obtained from the BGS data repository, item ID 187699 \cite{Harris2025}[Unrestricted access, Open Government Licence (OGL)].

\acknowledgments
We gratefully acknowledge sponsorship from the EPSRC (Grant Number EP/R513052/1) and BP. We acknowledge Computer Modelling Group (CMG) for providing access to IMEX.

\bibliography{Aux_files/ref.bib}

@article{Jackson2020,
author = {Jackson, Samuel J. and Krevor, Samuel},
doi = {10.1029/2020GL088616},
journal = {Geophysical Research Letters},
title = {{Small‐Scale Capillary Heterogeneity Linked to Rapid Plume Migration During CO\textsubscript{2} Storage}},
volume = {47},
number = {18},
pages = {e2020GL088616},
year = {2020}
}

@Misc{Harris2025,
  author    = {Harris, Catrin and Krevor, Samuel and Muggeridge, Ann H and Jackson, Samuel J},
  title     = {Medical X-ray CT data and model of rate-dependent capillary trapping in heterogeneous Bentheimer rock},
  year      = {2025},
  doi       = {10.5285/7896A0C1-0E8E-475A-89DB-7D844A61819B},
  publisher = {NERC EDS National Geoscience Data Centre},
}

@article{zhang2025,
title = {Capillary pinning in sedimentary rocks for CO2 storage: Mechanisms, terminology and State-of-the-Art},
journal = {International Journal of Greenhouse Gas Control},
volume = {144},
pages = {104385},
year = {2025},
issn = {1750-5836},
doi = {https://doi.org/10.1016/j.ijggc.2025.104385},
author = {Qin Zhang and Sebastian Geiger and Joep E.A. Storms and Denis V. Voskov and Matthew D. Jackson and Gary J. Hampson and Carl Jacquemyn and Allard W. Martinius}
}

@article{Dale1997,
author = {Dale, Magnar and Ekrann, Steinar and Mykkeltveit, Johannes and Virnovsky, George},
doi = {10.1023/A:1006536021302},
journal = {Transport in Porous Media},
title = {{Effective Relative Permeabilities and Capillary Pressure for One-Dimensional Heterogeneous Media}},
volume = {26},
number = {3},
pages = {229--260},
year = {1997}
}

@article{Duijn1995,
title = {The effect of capillary forces on immiscible two-phase flow in heterogeneous porous media},
volume = {21},
doi = {10.1007/BF00615335},
number = {1},
journal = {Transport in Porous Media},
author = {van Duijn, C. J. and Molenaar, J. and de Neef, M. J.},
year = {1995},
pages = {71--93}
}

@article{Saadatpoor2010,
author = {Saadatpoor, Ehsan and Bryant, Steven L. and Sepehrnoori, Kamy},
doi = {10.1007/s11242-009-9446-6},
journal = {Transport in Porous Media},
title = {{New Trapping Mechanism in Carbon Sequestration}},
volume = {82},
number = {1},
pages = {3--17},
year = {2010}
}

@book{BC,
author = {Brooks, R. H. and Corey, A. T.},
series = {Hydrology Papers No. 3},
title = {{Hydraulic properties of porous media}},
publisher = {Colorado State University},
year = {1964}
}

@article{Pini2013,
author = {Pini, Ronny and Benson, Sally M.},
doi = {10.1002/wrcr.20274},
journal = {Water Resources Research},
title = {{Simultaneous determination of capillary pressure and relative permeability curves from core-flooding experiments with various fluid pairs}},
volume = {49},
number = {6},
pages = {3516--3530},
year = {2013}
}

@article{Krevor2015,
author = {Krevor, Samuel and Blunt, Martin J. and Benson, Sally M. and Pentland, Christopher H. and Reynolds, Catriona and Al-Menhali, Ali and Niu, Ben},
doi = {10.1016/j.ijggc.2015.04.006},
journal = {International Journal of Greenhouse Gas Control},
title = {{Capillary trapping for geologic carbon dioxide storage - From pore scale physics to field scale implications}},
volume = {40},
pages = {221--237},
year = {2015}
}

@book{VanDuijnDeNeefM.1996,
title = {Self-similar profiles for capillary diffusion driven flow in heterogeneous porous media},
author = {van Duijn, C. J. and de Neef, M. J.},
year = {1996},
series = {{CWI report. AM-R; Vol. 9601}},
publisher = {Centrum voor Wiskunde en Informatica}
}

@article{Debbabi2017,
author = {Debbabi, Yacine and Jackson, Matthew D. and Hampson, Gary J. and Salinas, Pablo},
doi = {10.1007/s11242-017-0915-z},
journal = {Transport in Porous Media},
title = {{Capillary Heterogeneity Trapping and Crossflow in Layered Porous Media}},
volume = {120},
number = {1},
pages = {183--206},
year = {2017}
}

@phdthesis{Regnier2019,
author = {Regnier, Geraldine},
school = {Imperial College London},
title = {{Modelling of CO\textsubscript{2} plume dynamics at the lab-scale using an invasion percolation algorithm}},
year = {2019}
}

@article{Kortekaas1985,
author = {Kortekaas, T. F. M.},
doi = {10.2118/12112-PA},
journal = {Society of Petroleum Engineers Journal},
title = {{Water/Oil Displacement Characteristics in Crossbedded Reservoir Zones}},
volume = {25},
number = {6},
pages = {917--926},
year = {1985}
}

@article{Krevor2011, 
author = {Krevor, Samuel C.M. and Pini, Ronny and Li, Boxiao and Benson, Sally M.},
doi = {10.1029/2011GL048239},
journal = {Geophysical Research Letters},
title = {{Capillary heterogeneity trapping of CO\textsubscript{2} in a sandstone rock at reservoir conditions}},
volume = {38},
number = {15},
pages = {L15401},
year = {2011}
}

@article{Jackson2018, 
author = {Jackson, Samuel J. and Agada, Simeon and Reynolds, Catriona A. and Krevor, Samuel},
doi = {10.1029/2017WR022282},
journal = {Water Resources Research},
pages = {3139--3161},
title = {{Characterizing Drainage Multiphase Flow in Heterogeneous Sandstones}},
volume = {54},
number = {4},
year = {2018}
}

@article{Reynolds2018, 
author = {Reynolds, Catriona A. and Blunt, Martin J. and Krevor, Samuel},
doi = {10.1002/2017WR021651},
journal = {Water Resources Research},
title = {{Multiphase Flow Characteristics of Heterogeneous Rocks From CO\textsubscript{2} Storage Reservoirs in the United Kingdom}},
volume = {54},
number = {2},
pages = {729--745},
year = {2018}
}

@article{Jackson2020REV, 
author = {Jackson, S.J. and Lin, Q. and Krevor, S.},
doi = {10.1029/2019wr026396},
journal = {Water Resources Research},
title = {{Representative Elementary Volumes, Hysteresis and Heterogeneity in Multiphase Flow From the Pore to Continuum Scale}},
volume = {56},
number = {6},
pages = {e2019WR026396},
year = {2020}
}

@article{Pentland2011,
author = {Pentland, Christopher H. and El-Maghraby, Rehab and Iglauer, Stefan and Blunt, Martin J.},
doi = {10.1029/2011GL046683},
journal = {Geophysical Research Letters},
title = {{Measurements of the capillary trapping of super-critical carbon dioxide in Berea sandstone}},
volume = {38},
number = {6},
pages = {L06401},
year = {2011}
}

@article{VanDuijn2016,
author = {van Duijn, C. J. and Cao, X. and Pop, I. S.},
doi = {10.1007/s11242-015-0547-0},
journal = {Transport in Porous Media},
title = {{Two-Phase Flow in Porous Media: Dynamic Capillarity and Heterogeneous Media}},
volume = {114},
number = {2},
pages = {283--308},
year = {2016}
}

@article{Niu2015,
author = {Niu, Ben and Al‐Menhali, Ali and Krevor, Samuel C.},
doi = {10.1002/2014WR016441},
journal = {Water Resources Research},
pages = {2009--2029},
title = {{The impact of reservoir conditions on the residual trapping of carbon dioxide in Berea sandstone}},
volume = {51},
number = {4},
year = {2015}
}

@article{Ni2019,
author = {Ni, Hailun and Boon, Maartje and Garing, Charlotte and Benson, Sally M},
doi = {10.1016/j.ijggc.2019.04.024},
journal = {International Journal of Greenhouse Gas Control},
title = {{Predicting CO\textsubscript{2} residual trapping ability based on experimental petrophysical properties for different sandstone types}},
volume = {86},
pages = {158--176},
year = {2019}
}

@article{Li2015,
author = {Li, Boxiao and Benson, Sally M.},
doi = {10.1016/j.advwatres.2015.07.010},
journal = {Advances in Water Resources},
title = {{Influence of small-scale heterogeneity on upward CO\textsubscript{2} plume migration in storage aquifers}},
volume = {83},
pages = {389--404},
year = {2015}
}

@article{Krevor2012,
author = {Krevor, Samuel C.M. and Pini, Ronny and Zuo, Lin and Benson, Sally M.},
doi = {10.1029/2011WR010859},
journal = {Water Resources Research},
number = {2},
pages = {1--16},
title = {{Relative permeability and trapping of CO\textsubscript{2} and water in sandstone rocks at reservoir conditions}},
volume = {48},
year = {2012}
}

@article{Dance2016,
author = {Dance, T. and Paterson, L.},
doi = {10.1016/j.ijggc.2016.01.042},
journal = {International Journal of Greenhouse Gas Control},
pages = {210--220},
title = {{Observations of carbon dioxide saturation distribution and residual trapping using core analysis and repeat pulsed-neutron logging at the CO2CRC Otway site}},
volume = {47},
year = {2016}
}

@phdthesis{Reynoldsthesis2016,
author = {Reynolds, Catriona Anne},
school = {Imperial College London},
title = {{Two-phase flow behaviour and relative permeability between CO\textsubscript{2} and brine in sandstones at the pore and core scales}},
year = {2016}
}

@article{Ringrose1993,
author = {Ringrose, P S and Sorbie, K S and Corbett, P W M and Jensen, J L},
title = {{Immiscible flow behaviour in laminated and cross-bedded sandstones}},
doi = {10.1016/0920-4105(93)90071-L},
journal = {Journal of Petroleum Science and Engineering},
volume = {9},
number = {2},
pages = {103--124},
year = {1993}
}

@inproceedings{Kleppe1997,
author = {Kleppe, J. and Delaplace, P. and Lenormand, R. and Hamon, G. and Chaput, E.},
booktitle = {SPE Annual Technical Conference and Exhibition},
doi = {10.2118/38899-MS},
title = {{Representation of Capillary Pressure Hysteresis in Reservoir Simulation}},
address = {San Antonio, Texas},
pages = {SPE-38899-MS},
year = {1997}
}

@incollection{Krevor2020Chp,
title = {{Chapter 8: An Introduction to Subsurface CO\textsubscript{2} Storage}},
booktitle = {Carbon {Capture} and {Storage}},
publisher = {The Royal Society of Chemistry},
author = {Krevor, S. and Blunt, M. J. and Trusler, J. P. M. and De Simone, S.},
editor = {Reiner, David and Bui, Mai and Mac Dowell, Niall},
year = {2019},
doi = {10.1039/9781788012744-00238},
pages = {238--295}
}

@article{Bachu2013,
author = {Bachu, Stefan},
doi = {10.1016/j.egypro.2013.07.001},
journal = {Energy Procedia},
pages = {4428--4436},
title = {{Drainage and Imbibition CO\textsubscript{2}/Brine Relative Permeability Curves at in Situ Conditions for Sandstone Formations in Western Canada}},
volume = {37},
year = {2013}
}

@article{Al-Menhali2016,
author = {Al-Menhali, Ali S. and Krevor, Samuel},
doi = {10.1021/acs.est.5b05925},
journal = {Environmental Science and Technology},
title = {{Capillary Trapping of CO\textsubscript{2} in Oil Reservoirs: Observations in a Mixed-Wet Carbonate Rock}},
volume = {50},
number = {5},
pages = {2727--2734},
year = {2016}
}

@techreport{K43,
author = {{White Rose}},
title = {{K43: Field Development Report}},
year = {2016},
url = {https://assets.publishing.service.gov.uk/media/5a7ffcb3e5274a2e87db72d4/K43_Field_Development_Report.pdf}
}

@article{Harris2021,
author = {Harris, Catrin and Jackson, Samuel J. and Benham, Graham P. and Krevor, Samuel and Muggeridge, Ann H.},
doi = {10.1016/j.ijggc.2021.103511},
journal = {International Journal of Greenhouse Gas Control},
title = {{The impact of heterogeneity on the capillary trapping of CO\textsubscript{2} in the Captain Sandstone}},
volume = {112},
pages = {103511},
year = {2021}
}

@article{Pini2012,
author = {Pini, Ronny and Krevor, Samuel C.M. and Benson, Sally M.},
doi = {10.1016/j.advwatres.2011.12.007},
journal = {Advances in Water Resources},
pages = {48--59},
title = {{Capillary pressure and heterogeneity for the {CO\textsubscript{2}}/water system in sandstone rocks at reservoir conditions}},
volume = {38},
year = {2012}
}

@article{Ni2021,
author = {Ni, Hailun and M{\o}yner, Olav and Kurtev, Kuncho D. and Benson, Sally M.},
doi = {10.1016/j.advwatres.2021.103990},
journal = {Advances in Water Resources},
title = {{Quantifying CO\textsubscript{2} capillary heterogeneity trapping through macroscopic percolation simulation}},
volume = {155},
pages = {103990},
year = {2021}
}

@article{Mouche2010,
author = {Mouche, Emmanuel and Hayek, Mohamed and M{\"{u}}gler, Claude},
doi = {10.1016/j.advwatres.2010.07.005},
journal = {Advances in Water Resources},
title = {{Upscaling of CO\textsubscript{2} vertical migration through a periodic layered porous medium: The capillary-free and capillary-dominant cases}},
volume = {33},
number = {9},
pages = {1164--1175},
year = {2010}
}

@article{Herring2013,
author = {Herring, Anna L. and Harper, Elizabeth J. and Andersson, Linn{\'{e}}a and Sheppard, Adrian and Bay, Brian K. and Wildenschild, Dorthe},
doi = {10.1016/j.advwatres.2013.09.015},
journal = {Advances in Water Resources},
title = {{Effect of fluid topology on residual nonwetting phase trapping: Implications for geologic CO\textsubscript{2} sequestration}},
volume = {62},
pages = {47--58},
year = {2013}
}

@article{Bump2023,
author = {Bump, Alexander P. and Bakhshian, Sahar and Ni, Hailun and Hovorka, Susan D. and Olariu, Marianna I. and Dunlap, Dallas and Hosseini, Seyyed A. and Meckel, Timothy A.},
doi = {10.1016/j.ijggc.2023.103908},
journal = {International Journal of Greenhouse Gas Control},
pages = {{103908}},
title = {{Composite confining systems: Rethinking geologic seals for permanent CO\textsubscript{2} sequestration}},
volume = {126},
year = {2023}
}

@article{Iglauer2011,
author = {Iglauer, Stefan and W{\"{u}}lling, Wolfgang and Pentland, Christopher H. and Al-Mansoori, Saleh K. and Blunt, Martin J.},
doi = {10.2118/120960-PA},
journal = {SPE Journal},
title = {{Capillary-Trapping Capacity of Sandstones and Sandpacks}},
volume = {16},
number = {4},
pages = {778--783},
year = {2011}
}

@article{Wang2016, 
author = {Wang, Shibo and Tokunga, Tesu K. and Wan, Jiamin and Dong, Wenming and Kim, Yongman},
doi = {10.1002/2016WR018816.},
journal = {Water Resources Research},
volume = {52},
pages = {6671--6690},
title = {Capillary pressure-saturation relations in quartz and carbonate sands: Limitations for correlating capillary and wettability influences on air, oil, and supercritical CO$_2$ trapping},
year = {2016}
}

\end{document}